# Universal discontinuous percolation transition in the Earth's terrestrial topography

Shengjie Hu[1,2,3], Zhenlei Yang[2,3], Zipeng Wang[4], Sergio Andres Galindo Torres[2,3,#], Ling Li[2,3,#]

[1] College of Environmental and Resource Sciences, Zhejiang University; Hangzhou, Zhejiang province, 310058, China.

[2] School of Engineering, Westlake University; Hangzhou, Zhejiang province, 310030, China.

[3] Key Laboratory of Coastal Environment and Resources of Zhejiang Province, Westlake University; Hangzhou, Zhejiang province, 310030, China.

[4] School of Science, Westlake University; Hangzhou, Zhejiang province, 310030, China.

[#] Corresponding authors: Sergio Andres Galindo Torres; Ling Li
Email: s.torres@westlake.edu.cn; liling@westlake.edu.cn

**Abstract** Based on the Topographic Wetness Index (TWI), we studied the percolation process of water on the Earth's land surface and discovered a universal discontinuous phase transition across scales, with a critical TWI threshold of $0.671 \pm 0.054$. The discontinuity is attributed to the long-range correlation and directionality of the percolation process. Furthermore, the criticality is shown to extend from the critical point to a region corresponding to the Griffiths phase, where natural lake systems are found to develop, indicating the governess of self-organized criticality within the Earth system.

A long-standing puzzle for geoscientists, as well as physicists and mathematicians, is the ubiquity of scale-invariant fractal patterns on the Earth's land surface, manifested as the widely observed power law size distribution in various types of natural complex systems [1-2]. Topography, as the interface where these systems form and evolve, holds the key to understanding the common patterns and features. However, the specific connections between them remain unclear. Percolation theory in statistical physics provides a fundamental model for studying the connectivity of systems in which geometric phase transitions occur at critical points, accompanied by the emergence of patterns characterized by power law distributions [3]. Therefore, the application of percolation theory to the study of topography provides a new, quantitative way to explore the origin of fractal structures and critical phenomena on the Earth.

Saberi was the first to show a percolation transition in the Earth's topography associated with continental aggregation [4]. Subsequently, Fan et al. demonstrated that this phase transition is likely to be discontinuous due to the long-range correlation of elevation, with critical thresholds of 0.321 and 0.379 for land and ocean, respectively [5]. These results were obtained by analyzing the geometric features of islands and wetted clusters formed on the digital elevation map (DEM) as a "water level" (i.e., elevation cut) was gradually raised to flood the elevation surface [4, 5]. Because only a single elevation cut was applied to the entire study region (the whole Earth in both studies), the previous analyses were unable to account for both global and local elevation minima, especially among regions of different altitudes and landforms, such as coastal plains versus highland catchments. Consequently, the results mainly reflected the distinction between continents and oceans, leaving a gap in the percolation characteristics across different scales within the Earth's landmass. Moreover, water driven by gravity flows along the steepest gradient of the topographic relief network and accumulates in depressions [6], so that water percolation at the land surface is essentially a process with both directionality and long-range correlation. These essential features are however not captured by the single elevation-based method.

Here, we introduced the Topographic Wetness Index (TWI) instead of elevation to simulate the spatial distribution of water on the elevation surface to address the above limitations. With this approach, we studied the percolation process of water on topography in 14 regions of different sizes and topographic features at two observation scales of 30 m and 1 km. Our primary goal is to explore the universality in the percolation phase transition of the Earth's terrestrial topography. The second goal is to determine the relationship between the percolation properties of topography and the fractal patterns observed in natural complex systems, as exemplified by lakes. The results will advance our fundamental understanding of the complex Earth systems and provide insights particularly into the evolutionary mechanism of lake systems.

TWI, which combines upslope accumulation and downslope drainage, is a commonly used method to quantify the topographic control on the surface water runoff process in hydrology [6]. It is defined as $TWI = log_{10}(AS/tan\beta)$, where $AS$ is the upslope contributing area and $\beta$ is the local slope. The spatial distribution of TWI was determined separately for 11 hydro-climatic zones in China with topographic relief varying from 46 m to 234 m [7], and for the whole China, Asia and the globe using DEM data with 30 m and 1 km resolution, respectively (See Supplementary Material [8] for detailed information about the study regions and DEM data). The resolution of the DEM data indicates the ground size corresponding to a square cell on the data grid (e.g., a cell on a 30 m DEM has a ground area of 30×30 m$^2$); the lower its value, the more topographic detail is shown. Thus the change in resolution from high (30 m) to low (1 km) is similar to the coarse-graining process [9]. $AS$ was calculated via the D-8 algorithm [10], which determines the water flow from a given cell to a neighboring cell along the steepest downslope among its eight neighboring cells, and the $AS$ value represents how many other cells will drain into a given cell (Fig. 1a). It can be seen that TWI integrates global and local information about the topography of a region, and incorporates the directionality and long-rang correlation inherent in the water flow process (Fig. 1a). Therefore, it is a more appropriate parameter than elevation for studying the percolating behavior of topography.

According to the definition of TWI, the larger the TWI value is, the more likely water accumulates at the location. When water arrives, cells on the DEM with larger TWI values will be occupied first, followed by those with smaller TWI values, as water availability increases. Thus, the percolation model was constructed as follows: (1) In order to be comparable among regions, the TWI values were first normalized for each region, i.e., $TWI_N = \frac{TWI-TWI_{min}}{TWI_{max}-TWI_{min}}$, where $TWI_{min}$ and $TWI_{max}$ are the minimum and maximum values of TWI in the region, respectively. (2) We set different threshold $TWI_N$ values (denoted as $TWI\_p$) from high to low with an increment of $\Delta= -0.01$. For each threshold value, cells with $TWI_N$ greater than $TWI\_p$ are identified as wet cells occupied by water. Through this procedure, the configuration of wet cells was continuously determined for each $TWI\_p$, as shown in Figure 1b and movie S1. (3) Percolation properties were then be explored by analyzing how the geometric features of wetted clusters (a wetted cluster is a group of interconnected wet cells) change with $TWI\_p$.

Figure 2a and Figure S3 show that the area fraction ($A_w$, ratio of area of wetted clusters to the total area) and the average number ($N_w$, ratio of number of wetted clusters to the total number of DEM cells) of wetted clusters in each region increase with the decrease of $TWI\_p$. Two transitions occur at $TWI\_pc1$ and $TWI\_pc2$ ($TWI\_pc1 > TWI\_pc2$), with a relatively static region (the variations in the values of all analyzed percolation variables are less than 5% [8]) in the middle. The rapid increase of $A_w$ and $N_w$ below $TWI\_pc2$ indicates that the entire region is getting occupied by water (Fig. 1b), corresponding to extreme flooding events that would result in a spanning wetted cluster covering the whole region. In contrast, the transition at $TWI\_pc1$ has a non-trivial practical significance.

Systems on a network or a regular lattice are considered to be percolating when a giant cluster is present and the relative size of the largest cluster ($A_{max}$) is often used as the order parameter in percolation analysis [11]. Considering the comparability between study regions, we defined $A_{max} = \frac{S_1(TWI\_p)}{S_1(TWI\_pc1)}$, where $S_1(TWI\_pc1)$ and $S_1(TWI\_p)$ are the sizes (areas) of the largest wetted clusters formed at $TWI\_pc1$ and other $TWI\_p$ values in a region, respectively. In all regions, $A_{max}$ increases with the decrease of $TWI\_p$ and, and more importantly, shows an abrupt change at $TWI\_pc1$ (Fig. 2b). This suggests that $TWI\_pc1$ is a critical point where a discontinuous phase transition may occur, whereas no such discontinuity is observed at $TWI\_pc2$ (Fig. S4). The value of $TWI\_pc1$ is relatively constant across regions at $0.671 \pm 0.054$, while the value of $TWI\_pc2$ is more variable around $0.379 \pm 0.086$ (Fig. 2b and Table S2). Between $TWI\_pc1$ and $TWI\_pc2$, $A_{max}$ remains unchanged, indicating that criticality extends from a point to a region, reminiscent of the Griffiths phase [12].

We further assessed whether the phase transition observed at $TWI\_pc1$ is indeed discontinuous using the gap statistics-based approach [13]. According to Ref. [13], the phase transition is discontinuous if the largest gap, $\Delta g_{max} \coloneqq \max(S_1(TWI\_p + \Delta) - S_1(TWI\_p))$, is macroscopic (extensive) to the system size ($S_s$, area of the study region). It is strongly discontinuous if $\lim_{S_s \to \infty} \frac{\Delta g_{max}}{S_s} > 0$; otherwise it is weakly discontinuous. Natural systems are of finite sizes and in our case the upper size limit is the global land area. The value of $\frac{\Delta g_{max}}{S_s}$ of the globe is 0.002431, a small value close to 0. As the observation scale used in China's 11 hydro-climatic zones (30 m) is smaller than that in the other regions (1 km), the DEM data from the former correspond to a smaller ground area but a larger number of cells (Fig. 2c). Given that the number of cells ($S_N$) is another proxy for the system size in percolation on a regular lattice [3, 13], we analyzed the variation of $\frac{\Delta g_{max}}{S_s}$ value with increasing $S_N$ and found a decreasing trend, with a minimum value of 0.000011 (Fig. 2c and Table S2). Based on these results, we inferred that the percolation transition occurring at $TWI\_pc1$ is (weakly) discontinuous.

Although the phase transition found here and in Ref. [5] are both discontinuous, the methods used are quite different, with the control parameters being $TWI\_p$ and occupation probability, respectively. With $TWI\_p$, we showed that the discontinuous phase transition is universal over the Earth's land surface, with a critical threshold of $0.671 \pm 0.054$, independent of the observation scale and the details of the study region. If the occupation probability from Ref. [5], i.e., $A_w$ here, is used as the control parameter, $A_{max}$ still shows abrupt changes, but the corresponding critical thresholds of $A_w$ range from 0.0006 to 0.3573 across the different regions (Fig. 2d and Fig. S6), influenced by the regional topographic features and DEM resolution (see Supplementary Material [8] for detailed discussion). These results suggest that the percolation process of water on the (gravitational) potential energy surface, dominated by the water flow mechanism, is well captured by the TWI method, which led to the discovery of the universal percolation phase transition in topography.

The discontinuity observed in Ref. [5] was attributed to the long-range correlation of elevation. Long-range correlation also plays an important role in the discontinuous phase transition revealed here, but it originates from the water movement process, as reflected by the parameter $AS$ in TWI (Fig. 1a). Directionality is another contributing factor [14], as evidenced by the different behavior of the largest wetted cluster size ($S_1(TWI\_p)$) and the average wetted cluster size ($S_a(TWI\_p) = \frac{S_{Tot}(TWI\_p)}{N(TWI\_p)}$, where $S_{Tot}(TWI\_p)$ and $N(TWI\_p)$ are the total area and number of wetted clusters formed at each $TWI\_p$, respectively) above and below the critical region (Fig. 2e and Fig. S5). In the subcritical region ($TWI\_p > TWI\_pc1$), both $S_1(TWI\_p)$ and $S_a(TWI\_p)$ increase as $TWI\_p$ decreases, indicating that the water has flowed to and accumulated at certain locations, i.e., the percolation is directed. On the contrary, in the supercritical region ($TWI\_p < TWI\_pc2$), $S_1(TWI\_p)$ either remains unchanged or shows jumps, while $S_a(TWI\_p)$ decreases rapidly. This demonstrates that the water tends to be distributed in different locations with more wetted clusters formed and the percolation becomes undirected. Therefore, the directionality and long-range correlation in the water distribution process explain the discontinuity of the percolation transition observed at $TWI\_pc1$ in all study regions. According to Ref. [13], whether the discontinuity is strong or weak is related to the cluster joining dynamics (see Supplementary Material [8] for details). If the largest cluster is allowed to grow directly from the largest one formed in the previous step, the percolation process will exhibit a weak discontinuous phase transition. We found that this is what happened at $TWI\_pc1$ in most of the study regions (Table S1).

Another percolation variable, the mean cluster size (χ), was also measured. It is defined as $\chi = \frac{\sum_s s^2 n_s(TWI\_p)}{\sum_s s n_s(TWI\_p)}$, where $n_s(TWI\_p)$ is the average number of wetted clusters of size $s$ at $TWI\_p$, and the largest wetted cluster was excluded in the calculation below $TWI\_pc2$ [3-5]. The value of χ and its variation (i.e., $C_\chi = \frac{\chi(TWI\_p+\Delta)-\chi(TWI\_p)}{\chi(TWI\_p)}$) for the studied regions are shown in Figure 2f and Figure S9. It can be seen that χ undergoes a sharp increase at $TWI\_pc1$, indicating that χ tends to diverge at the critical point [4]. The divergence of χ signals that the system becomes scale-invariant with power law distributed clusters [4]. Since natural lakes result from water occupying the land surface and their size follows a power law distribution when they are predominantly affected by topography [15], we hypothesized that the development of natural lake systems is associated with the discontinuous phase transition at $TWI\_pc1$.

To test this hypothesis, we compared wetted clusters formed at each $TWI\_p$ with data of natural lakes from the HydroLAKES database [16] for China's 11 hydro-climatic zones. We found that the size of the largest lake in each zone is close to the size of the largest wetted cluster formed at $TWI\_pc1$, while the size of the largest wetted cluster generated in the previous step ($TWI\_p =$ $TWI\_pc1 + 0.01$) can be an order of magnitude smaller (Table 1). This demonstrates that the largest lake dose not form until the discontinuous percolation transition occurs. The spatial consistency ($S\_C$, proportion of lakes that spatially overlap with wetted clusters to the total number of lakes in a study zone) between simulated wetted clusters and lakes also significantly improves at $TWI\_pc1$ compared to the previous step (Fig. 3a and Fig. S10). Furthermore, the size distributions of natural lakes and wetted clusters formed at $TWI\_pc1$ both follow the power law with similar values of power exponents (Fig. 3b and 3c, Fig. S11 and Table 1). In contrast, the wetted clusters formed one step before could be better fitted with a stretched-exponential distribution (Fig. 3d and Fig. S11), especially in the Qiangtang Basin (QB), where the human interference is almost negligible [7]. These features and behaviors remain unchanged in $TWI\_pc2 \leq TWI\_p \leq TWI\_pc1$, indicating that the extended critical region corresponds to a Griffiths phase [12]. Combining all these results, we concluded that lake systems develop in the critical region of the discontinuous percolation transition of topography, providing the first evidence for the governess of self-organized criticality in this complex system. Since the Griffiths phase and stretching criticality are also found in biological systems, such as brain networks [12], similar behaviors found in the lake system, an abiotic system, seem to support the conjecture that complex systems share common dynamic principles [2].

Despite self-organized criticality is considered as a general principle for the state evolution of dynamical systems, why systems operate at the critical state remains an open question [2, 12]. It has been hypothesized that a system operating at the criticality has functional advantages, as the result of an evolutionary/adaptive process [12]. The effective potential, $EP = -log(Prob(s))$, where $Prob(s)$ is the probability distribution of cluster size, measures the stability of a macroscopic state [17-18]. A more stable state seems to have a higher curvature in the change of $EP$ and the most likely state is the one with the minimum $EP$ [17-18]. We calculated $EP$ for the wetted clusters formed at each step of the simulation, and found that in most regions $EP$ decreases significantly at $TWI\_pc1$ and maintains a low value throughout the critical region but increases again below $TWI\_pc2$ (Fig. 4 and Fig. S12). This suggests that criticality may arise from the quest for stability. Other potential evolutionary benefits of self-organized criticality to the lake system, especially in terms of the feedback to environmental changes, need more research to be determined.

To summarize, we have revealed a universal discontinuous percolation transition in the Earth's terrestrial topography and shown that it is closely related to the development of natural lake systems. The universality results from the inclusion of the water flow mechanism on the elevation surface in the percolation analysis, thus the approach applied here can be informative for studying percolation processes on other energy landscapes in flow representation. Since universality is a pillar of modern critical phenomena, understanding the profound physical and mathematical principles underlying the universal phenomenon found here should be the focus of future work. In addition to revealing the controlling role of the self-organized criticality in lake systems, our results warn that lake systems can be driven out of the critical region and suddenly enter unstable states. The widespread decline in global lake water storage recently reported [19] may be an indication that lake systems are approaching the critical point of discontinuous percolation transition due to a combination of human disturbance and climate change. Such a shift will pose a major threat to lakes and adjacent eco-hydrological systems, as well as to water security for the human society. Therefore, how the state of lake systems responds to environmental change should also be an important research question for future work.

**Acknowledgments**

The authors acknowledge funding support from National Natural Science Foundation of China (Grant No.41976162).

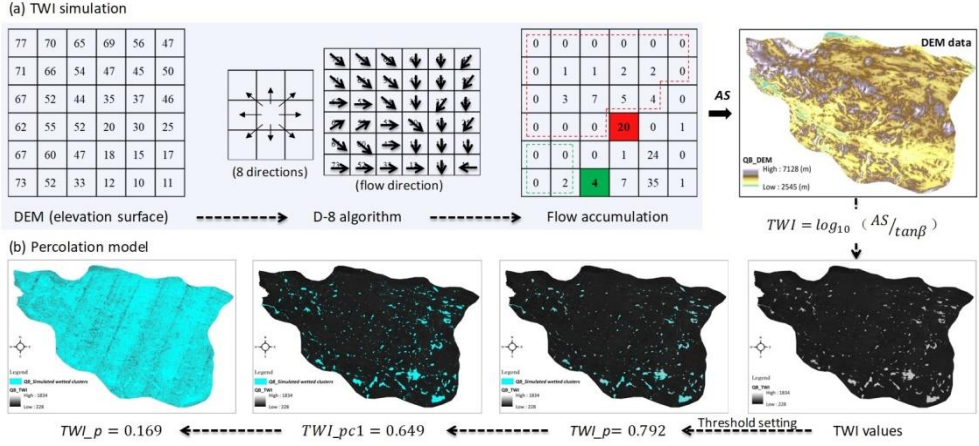

**Fig. 1.** Simulation of the water percolation process on the elevation surface using the Topographic Wetness Index (TWI). (a) TWI simulation: values on the flow accumulation map represent the number of cells that will drain to each cell. For example, the contributing area of the red and green colored cells consists of the cells within the red and green dashed lines, respectively; (b) Percolation model: wetted clusters formed with TWI threshold ($TWI\_p$) varying from high to low. The QB (Qiangtang Basin) hydro-climatic zone is used as an example to illustrate the simulation process.

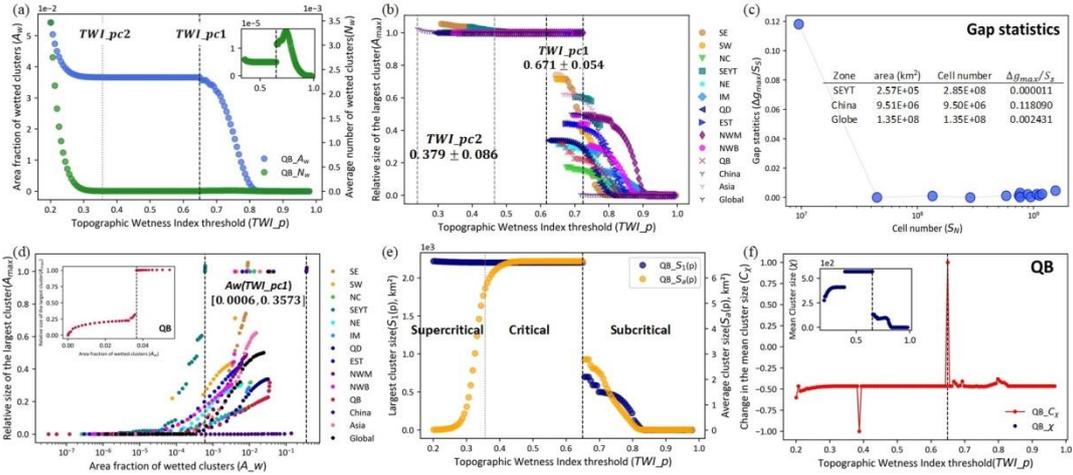

**Fig. 2.** Results of percolation analysis. (a) Increasing trend of the area fraction ($A_w$) and the average number ($N_w$) of wetted clusters with decreasing $TWI\_p$ value: two transition points ($TWI\_pc1$ and $TWI\_pc2$) with a static region in between; (b) The order parameter, the relative size of the largest wetted cluster ($A_{max}$), changes abruptly at $TWI\_pc1 = 0.671 \pm 0.054$ in all study regions, indicating that a universal discontinuous phase transition may occur there; (c) Gap statistics demonstrates that the phase transition at $TWI\_pc1$ is most likely to be weakly discontinuous; (d) When using $A_w$ as the control parameter, the critical threshold for the discontinuous phase transition varies considerably among regions; (e) Different behaviors of the largest cluster size ($S_1(TWI\_p)$) and the average cluster size ($S_a(TWI\_p)$) in the subcritical and supercritical regions suggest a change in the directionality of the percolation process; (f) The sharp increase of mean cluster size ($\chi$) at $TWI\_pc1$ implies a divergent tendency (the value of $C_\chi$ has been normalized). See Supplementary Material [8] for results of other study regions.

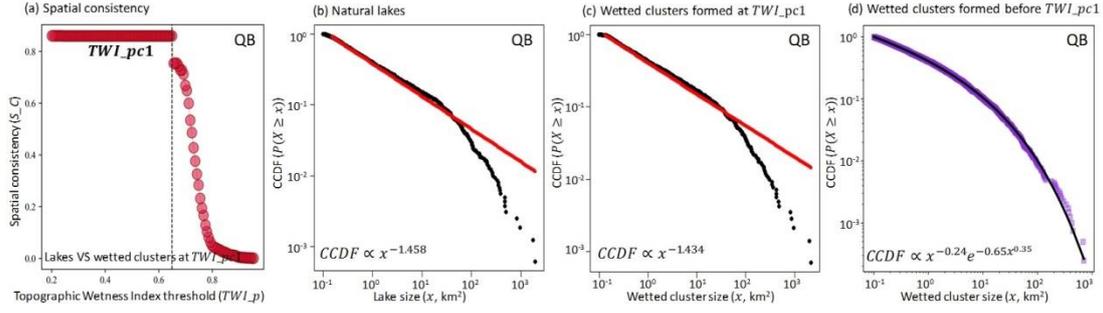

**Fig. 3.** Comparison between natural lakes and simulated wetted clusters in the QB zone. (a) Spatial consistency between lakes and wetted clusters has significantly improved at $TWI\_pc1$; (b)-(c) Power law size distributions of lakes and wetted clusters formed at $TWI\_pc1$; (d) The size of wetted clusters formed one step before $TWI\_pc1$ follows the stretched-exponential distribution. See Supplementary Material [8] for results of other zones.

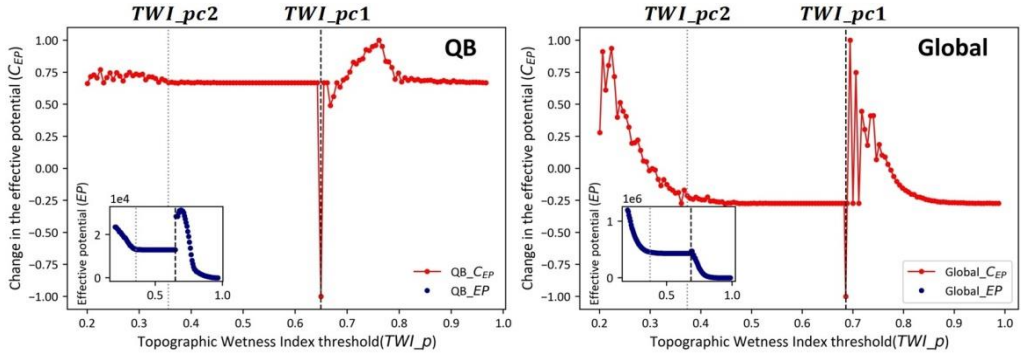

**Fig. 4.** Effective potential ($EP$) and its variation ($C_{EP}$) in the QB zone (left) and the globe (right): $EP$ of the wetted clusters decreases significantly at $TWI\_pc1$ and then maintains in a relatively low value between $TWI\_pc1$ and $TWI\_pc2$, suggesting that the critical region is more stable. The value of $C_{EP}$ has been normalized and results for other regions are in Supplementary Material [8].

**Table 1.** Comparison between natural lakes and simulated wetted clusters formed at $TWI\_pc1$ and its previous step ($TWI\_pc1 - \Delta$) in the hydro-climatic zones. $S_{L\_L}$ is the area of the largest lake; $S_1(TWI\_pc1)$ and $S_1(TWI\_pc1 - \Delta)$ represents the areas of the largest wetted clusters formed at $TWI\_pc1$ and $TWI\_pc1 - \Delta$; $\alpha_L$ and $\alpha_{TWI\_pc1}$ denote the exponents of power law size distributions of lakes and wetted clusters formed at $TWI\_pc1$, respectively. The unit of area is km$^2$.

| Zone | $S_{L\_L}$ | $S_1(TWI\_pc1)$ | $S_1(TWI\_pc1 - \Delta)$ | $\alpha_L$ | $\alpha_{TWI\_pc1}$ |
|---|---|---|---|---|---|
| SE | 4.246E+02 | 3.893E+02 | 3.090E+02 | 2.469 | 2.329 |
| SW | 2.983E+02 | 2.815E+02 | 2.010E+02 | 2.258 | 2.196 |
| NC | 1.374E+03 | 1.593E+03 | 5.083E+02 | 2.187 | 1.873 |
| SEYT | 4.280E+00 | 3.366E+00 | 1.996E+00 | 2.177 | 2.394 |
| NE | 1.044E+03 | 1.182E+03 | 3.665E+02 | 2.098 | 1.959 |
| IM | 2.121E+03 | 2.020E+03 | 5.797E+02 | 1.992 | 1.962 |
| QD | 2.398E+03 | 2.425E+03 | 8.197E+02 | 1.867 | 1.866 |
| EST | 6.178E+02 | 6.085E+02 | 2.680E+02 | 1.781 | 1.715 |
| NWM | 4.267E+03 | 4.112E+03 | 2.015E+03 | 1.668 | 1.620 |
| NWB | 9.618E+02 | 9.460E+02 | 4.661E+02 | 1.667 | 1.668 |
| QB | 1.964E+03 | 2.202E+03 | 6.989E+02 | 1.458 | 1.434 |